\documentclass [11pt, one column, a4paper] {article}

\usepackage[dvips]{epsfig}
\usepackage{amsmath}
\usepackage{amssymb}
\usepackage{graphicx}
\usepackage{bm}
\usepackage{epstopdf}
\usepackage[unicode=true,
 bookmarks=true,bookmarksnumbered=false,bookmarksopen=false,
 breaklinks=false,pdfborder={0 0 1},backref=false,colorlinks=false]
 {hyperref}
\usepackage{color}

\title{Higher order derivatives extension of Maxwell-Chern-Simons electrodynamics in the presence of field sources and material boundaries}
\author{L.H.C. Borges$^{1}$\thanks{email: luizhenriqueunifei@yahoo.com.br}, F.A. Barone$^{2}$\thanks{email: fbarone@unifei.edu.br}, H.L. Oliveira$^{3}$\thanks{email: helderluiz10@gmail.com}\\
{\small $^{1}$Universidade Federal do ABC, Centro de Ci\^encias Naturais e Humanas,}\\
{\small Rua Santa Ad\'elia, 166, 09210-170, Santo Andr\'e, SP, Brazil}\\
{\small $^{2}$IFQ - Universidade Federal de Itajub\'a, Av. BPS 1303, Pinheirinho,}\\
{\small Caixa Postal 50, 37500-903, Itajub\'a, MG, Brazil}\\
{\small $^{3}$UNESP - Campus de Guaratinguet\'a - DFQ, Avenida Dr. Ariberto Pereira da Cunha 333,}\\
{\small CEP 12516-410, Guaratinguet\'a, SP, Brazil}}

\date {}

\begin {document}

\baselineskip=12pt

\maketitle

\begin{abstract}
In this paper we consider some new classical effects obtained for a planar electrodynamics with the presence of a higher order derivatives term. The model can be interpreted as a kind of extension for the $3d$ Maxwell-Chern-Simons electrodynamics with higher order derivatives. We consider setups with stationary field sources describing point-like charges and Dirac points. We also investigate this model with the presence of a conducting line (in $3d$ dimensions, it is the equivalent of a conducting plate in $4d$ dimensions). In this case we calculate the propagator for the gauge field and the interaction force between the conducting line and a point-like charge, as well as the force between the conducting line and a Dirac point, the source for vortex field solutions. It is shown that the image method is not valid in any case. We also compare the obtained results along the paper with the corresponding ones obtained with the standard Maxwell-Chern-Simons electrodynamics \cite{LHCHLOFAB}. 
\end{abstract}

\section{Introduction}
\label{introduction}

Field theories in (2+1) dimensions have been intensely investigated in the literature and, maybe, the main  reason for this relies on the fact that such kind of theory exhibits many interesting aspects, at classical and quantum levels, due to the odd space-time dimensionality. In this scenario, maybe the most popular planar gauge field theory is the so called Maxwell-Chern-Simons electrodynamics \cite{Dunne,MCS1,MCS2}, whose features were considered in a variety of contexts. We can mention, for instance, the planar massive quantum electrodynamics (QED3) \cite{QED1,QED2,QED3,QED4,QED5}, its relevance in condensed matter systems (see, for instance, Ref. \cite{CMP} and references therein), the planar noncommutative electrodynamics \cite{NC1,NC2,NC3,NC4}, and the planar electrodynamics with boundary conditions \cite{Casimir1,Casimir2,Casimir3,Casimir4,Casimir5,Casimir6,Planar1,Planar2,Planar3,Planar4,Planar5}, among others.

In a recent work \cite{LHCHLOFAB} it was investigated some physical phenomena which emerge due to the presence of stationary field sources and a perfectly conducting line in the Maxwell-Chern-Simons electrodynamics. It is worth mentioning that, once we have a planar theory, the equivalent to a conducting plate is just a conducting line. The results were compared with the equivalent ones obtained in the standard (2+1)-dimensional Maxwell electrodynamics. 

Besides, models which exhibit higher order derivatives, including the planar ones, have been also studied in the literature from time to time, mainly because the inclusion of higher order derivatives improve renormalization properties that tame ultraviolet divergences in field theories, at classical and quantum level. In this context, we highlight the higher order derivative extension of the Chern-Simons electrodynamics proposed in Ref. \cite{hoch}, whose stability, causality and unitarity properties and their conservation laws have been investigated in Ref. \cite{DIS,VDS,PetrovSch}. In addition, this model was considered in the noncommutative context \cite{Bufalo} and its Hamilton-Jacobi analysis was carried out thoroughly \cite{EscalanteAlberto}.

The higher order derivative extension of the Chern-Simons term can be obtained from the standard QED in (2+1) dimensions, as an one-loop quantum correction when we integrate out in the fermionic field \cite{PetrovQED3}.

Recently, the higher order derivative extension of the Chern-Simons term, proposed in \cite{hoch}, has been considered to describe planar crystalline insulators \cite{ScRep21998}. Besides, a simplified version of this term (with the presence of the Laplacian instead of the d'Alembertian) has been used to describe viscosity in quantum Hall fluids \cite{PRB155425}.  

There is still a gap in the literature regarding the physical phenomena which emerge in the presence of field sources and a conducting line in the higher order derivative extension of the Maxwell-Chern-Simons electrodynamics. So, in this paper, we make a contribution to that extent, by considering the model discussed in the work of reference \cite{hoch} with the presence of stationary point-like sources and a perfectly conducting line. Specifically, section \ref{higherMCS1a} is devoted to an analysis of the free model propagator (without the presence of conductors). In section \ref{higherMCS1b} we consider effects due to the presence of point-like stationary charges and Dirac points. In section \ref{higherMCS2} we compute the propagator for the gauge field in the presence of a conducting line. In section \ref{higherMCS3} we obtain the interaction force between the conducting line and a point-like charge. We also compare the interaction force obtained in the free theory (theory without the conducting line) and we check that the image method is not valid for the higher order derivative theory considered in this work. In section \ref{higherMCS4} we make a similar analysis to the one of the section \ref{higherMCS3}, but for the so called Dirac point instead of charges, and we verify that the image method is not valid. We also compare the obtained results along the paper with the ones obtained for the Maxwell-Chern-Simons electrodynamics \cite{LHCHLOFAB}. Section \ref{conclusoes} is dedicated to our final remarks and conclusions.

Along the paper we shall deal with a model in $2+1$ dimensions in a Minkowski spacetime with diagonal metric $(+, -, -)$. The Levi-Civita tensor is denoted by $\epsilon^{\mu\nu\lambda}$ with $\epsilon^{012}=1$.

\section{\label{higherMCS1a} The model and its propagator}

In this section we investigate some interactions between stationary point-like field sources which arises from the higher order derivative extension of the $3d$ Maxwell-Chern-Simons electrodynamics, whose Lagrangian density is \cite{hoch}
\begin{equation}
\label{lchho}
{\cal L}= -\frac{1}{4}F_{\mu\nu}F^{\mu\nu}-\frac{1}{2\xi}\left(\partial_{\mu}A^{\mu}\right)^{2}+
\frac{1}{2m}\ \epsilon^{\mu\nu\lambda}\left(\Box A_{\mu}\right)\left(\partial_{\nu}A_{\lambda}\right)-J^{\mu}A_{\mu} \ ,
\end{equation}
where $A^{\mu}$ is the gauge field, $F^{\mu\nu}=\partial^{\mu}A^{\nu}-\partial^{\nu}A^{\mu}$ is the field strength, $J^{\mu}$ is the external source, $\xi$ is a gauge fixing parameter and $m>0$ is a positive parameter with dimension of mass. 

The model (\ref{lchho}) can be rewrite in the following way 
\begin{eqnarray}
\label{lhomcsoperator}
{\cal L}\rightarrow\frac{1}{2}A_{\mu}{\cal O}^{\mu\nu}A_{\nu} \ ,
\end{eqnarray}
where we defined the differential operator
\begin{eqnarray}
\label{diffopehomcs}
{\cal O}^{\mu\nu}=\Box\eta^{\mu\nu}-\left(1-\frac{1}{\xi}\right)\partial^{\mu}\partial^{\nu}-\frac{1}{m}\epsilon^{\mu\nu\rho}\Box\partial_{\rho} \ .
\end{eqnarray}

The propagator $D^{\mu\nu}\left(x,y\right)$ is the inverse of the operator ${\cal O}^{\mu\nu}$, as follows
\begin{eqnarray}
\label{inveseoperhomcs}
{\cal O}^{\mu\nu}D_{\nu\lambda}\left(x,y\right)=\eta^{\mu}_{ \ \lambda}\delta^{3}\left(x-y\right) \ .
\end{eqnarray}

By using standard field theory methods, we can show that the propagator in the Feynman gauge, $\xi=1$, is given by 
\begin{equation}
\label{propcho}
D^{\mu\nu}\left(x,y\right)=\int\frac{d^{3}p}{(2\pi)^{3}}\left(\frac{1}{p^{2}-m^{2}}-\frac{1}{p^{2}}\right)
\left(\eta^{\mu\nu}-\frac{p^{\mu}p^{\nu}}{m^{2}}-\frac{i}{m}\epsilon^{\mu\nu\lambda}p_{\lambda}\right)e^{-ip\cdot(x-y)} \ .
\end{equation}

The propagator (\ref{propcho}) reveals that the parameter $m$ takes on multiple roles. From the first parenthesis on the right hand side of (\ref{propcho}) we can see the presence a massive pole for $p^{2}=m^{2}$ and a massless pole. So, the model (\ref{lchho}) proposed in \cite{hoch} exhibits two kinds of field modes, ones with mass and others masslesss. It is a similar situation to the one found in the Lee-Wick electrodynamics \cite{DiracString4}. Besides, it is a different feature in comparison with the Maxwell-Chern-Simons electrodynamics, which exhibits just massive field modes.

The presence of a Levi-Civita (pesudo-)tensor in the third term on the right hand side of the lagrangian (\ref{lchho}) brings about that the model (\ref{lchho}) exhibits properties in common with the Maxwell-Chern-Simons electrodynamics, whose Lagrangian and propagator are given by \cite{LHCHLOFAB}
\begin{eqnarray}
\label{LDMCS}
{\cal L}_{\textrm{MCS}}&=&-\frac{1}{4}F_{\mu\nu}F^{\mu\nu}-\frac{1}{2\xi}\left(\partial_{\mu}A^{\mu}\right)^{2}+
\frac{m}{2}\epsilon^{\mu\nu\lambda}A_{\mu}\partial_{\nu}A_{\lambda}\cr\cr
D^{\mu\nu}_{(\textrm{MCS})}\left(x,y\right)&=&-\int\frac{d^{3}p}{(2\pi)^{3}}\frac{1}{p^{2}-m^{2}}
\left(\eta^{\mu\nu}-m^{2}\frac{p^{\mu}p^{\nu}}{p^{4}}+\frac{im}{p^{2}}\epsilon^{\mu\nu\lambda}p_{\lambda}\right)e^{-ip\cdot(x-y)} \ ,
\end{eqnarray}
in the gauge $\xi=1$. The sub-index $MCS$ means Maxwell-Chern-Simons.

The terms with Levi-Civita tensor in the propagators (\ref{propcho}) and (\ref{LDMCS}) give rise to a wide range of physical phenomena related to vortex field solutions in both theories, whose intensities are controlled by the parameter $m$. This point shall be explored in section (\ref{higherMCS1b}).

\section{\label{higherMCS1b} Interaction between external sources}

The theory (\ref{lchho}) is quadratic in the field variables $A^{\mu}$, so it can be shown that the contribution of the source $J^{\mu}(x)$ to the vacuum energy of the system is given by \cite{LHCHLOFAB,Zee,BaroneHidalgo1,BaroneHidalgo2} 
\begin{equation}
\label{energyS}
E=\frac{1}{2T}\int\int d^{3}x\ d^{3}y J^{\mu}(x)D_{\mu\nu}(x,y)J^{\nu}(y)\ ,
\end{equation}
where $T$ is the time variable and it is implicit the limit $T\rightarrow\infty$.

We start by considering the interaction between two point-like charges. This configuration is described by the external source 
\begin{eqnarray}
\label{corre1Em}
J^{CC}_{\mu}({\bf x})=\sigma_{1}\eta_{\ \mu}^{0}\delta^{2}\left({\bf x}-{\bf a}_ {1}\right)+\sigma_{2}\eta_{\ \mu}^{0}\delta^{2}\left({\bf x}-{\bf a}_ {2}\right) \ ,
\end{eqnarray}
where the location of the charges are specified by the spatial vectors ${\bf a}_ {1}$ and ${\bf a}_ {2}$ and the parameters $\sigma_{1}$ and $\sigma_{2}$ stand for the electric charges (in $2+1$ dimensions). The super-index $CC$ means that we have the interaction  between two point-like charges. 

Substituting (\ref{propcho}) and (\ref{corre1Em}) in (\ref{energyS}), discarding the self-interacting contributions,  performing the integrals in the following order,  $d^{2}{\bf x}$, $d^{2}{\bf y}$, $dx^{0}$, introducing the Fourier representation for the Dirac delta function, integrating out in the momenta $dp^{0}$ and identifying the time interval as $T=\int dy^{0}$, we arrive at
\begin{eqnarray}
\label{eeeenergy}
E^{CC}=\sigma_{1}\sigma_{2}\left(\int\frac{d^{2}{\bf p}}{\left(2\pi\right)^{2}}\frac{e^{i{\bf p}\cdot{\bf a}}}{{\bf p}^{2}}-\int\frac{d^{2}{\bf p}}{\left(2\pi\right)^{2}}\frac{e^{i{\bf p}\cdot{\bf a}}}{{\bf p}^{2}+m^{2}}\right) \ ,
\end{eqnarray}
where  ${\bf{a}}={\bf {a}}_{1}-{\bf {a}}_{2}$ stands for the distance between the two electric charges.

Notice that the energy (\ref{eeeenergy}) splits into two contributions. The first one comes from the massless sector of the model and does not involve the parameter $m$. The second one comes from the massive sector.  

For the second contribution we use the fact that \cite{BaroneHidalgo1}
\begin{eqnarray}
\label{int4EM}
\int\frac{d^{2}{\bf p}}{(2\pi)^{2}}\frac{e^{i{\bf p}\cdot{\bf a}}}
{{\bf p}^2+m^2}=\frac{1}{2\pi}K_{0}(ma) \ ,
\end{eqnarray}
where $a=\mid\bf{a}\mid$, and $K$ stands for the K-Bessel function \cite{Arfken}.

For the first contribution we insert a regulator parameter, $\mu$, with mass dimension, as follows \cite{LHCHLOFAB,BaroneHidalgo1,DiracString1}
\begin{eqnarray}
\label{INTD2}
\int\frac{d^{2}{\bf p}}{(2\pi)^{2}}\frac{e^{i{\bf p}\cdot{\bf a}}}{{\bf p}^2}&\to&\lim_{\mu\rightarrow0}
\int\frac{d^{2}{\bf p}}{(2\pi)^{2}}\frac{e^{i{\bf p}\cdot{\bf a}}}{{\bf p}^2+\mu^{2}}
=\frac{1}{2\pi}\lim_{\mu\rightarrow0}\left[K_{0}\left(\mu a\right)\right]=
-\frac{1}{2\pi}\lim_{\mu\rightarrow0}\left[\ln\left(\frac{\mu a}{2}\right)+\gamma\right]\nonumber\\
&=&-\frac{1}{2\pi}\lim_{\mu\rightarrow0}\left[\ln\left(\frac{\mu a}{2}\right)+\gamma+\ln(\mu a_{0})-
\ln(\mu a_{0})\right]\nonumber\\
&=&-\frac{1}{2\pi}\left[\ln\left(\frac{a}{a_{0}}\right)+\gamma-\ln 2+\lim_{\mu\rightarrow0}
\ln(\mu a_{0})\right]\nonumber\\
&\to&-\frac{1}{2\pi}\ln\left(\frac{a}{a_{0}}\right) \ ,
\end{eqnarray}
where in the first line we used the expansion $K_{0}(\mu a)\stackrel{\mu\rightarrow0}{\rightarrow}
-\ln\left(\mu a/2\right)-\gamma$ ($\gamma$ stands for the Euler constant) and in the second line, we added and subtracted the quantity $\ln\left(\mu a_{0}\right)$, where $a_{0}$ is an arbitrary constant with dimension of length. In the third line we discarded the $a$-independent terms, which does not contribute to the interaction energy between the charges and, so, to the force between them.

Inserting (\ref{int4EM}) and (\ref{INTD2}) in (\ref{eeeenergy}), the interaction energy between the stationary charges becomes
\begin{equation}
\label{eeeenergy2}
E^{CC}=-\frac{\sigma_{1}\sigma_{2}}{2\pi}\left[\ln\left(\frac{a}{a_{0}}\right)+K_{0}\left(ma\right)\right]\ ,
\end{equation}
and the interaction force reads
\begin{equation}
\label{forhoch}
F^{CC}=-\frac{dE^{CC}}{da}= \frac{\sigma_{1}\sigma_{2}}{2\pi a}\left[1-\left(ma\right)K_{1}\left(ma\right)\right] \ .
\end{equation}

In Eq. (\ref{forhoch}) the first term between brackets on the right hand side is the well-known $2+1$ dimensional Coulombian interaction. The $m$-dependent contribution is similar to that one obtained in Maxwell-Chern-Simons electrodynamics for the interaction between two point-like charges, but with an overall minus signal \cite{LHCHLOFAB}. This fact can be understood by considering the propagators of both theories. The relevant terms of the propagators of each theory, in this case, can be taken from (\ref{propcho}) and (\ref{LDMCS}) and are given by
\begin{eqnarray}
\label{propart1}
D_{CC}^{\mu\nu}\left(p\right)\sim\left(\frac{1}{p^{2}-m^{2}}-\frac{1}{p^{2}}\right)\eta^{\mu\nu}, \ \ 
D_{CC(\textrm{MCS})}^{\mu\nu}\left(p\right)\sim-\frac{1}{p^{2}-m^{2}}\eta^{\mu\nu} \ .
\end{eqnarray}
By comparing the above expressions, it is evident to see where the overall minus signal differentiating the $m$-dependent contributions in both theories comes from. The term in brackets in (\ref{forhoch}) is always non-negative and goes to 1 for large values of $ma$, so the force is repulsive for charges with the same signal and exhibits a Coulombian behavior (in $2+1$ dimensions) for large $ma$ values.

For small values for the distance $a$, the force (\ref{forhoch}) goes to zero, as usual for field theories with higher order derivatives.

In the next example we study the interaction energy between a point-like charge and a Dirac point. Such a system is composed by the following external field source
\begin{eqnarray}
\label{corre112Em}
J^{CD}_{\mu}({\bf x})=\sigma\eta_{\ \mu}^{0}\delta^{2}\left({\bf x}-{\bf a}_ {1}\right)+
J_{\mu(D)}\left({\bf x}\right) \ ,
\end{eqnarray}
where the first term stands for the external field source produce by the point-like charge placed at position ${{\bf {a}}_{1}}$ and the second one is the source produced by the Dirac point. The super-index $CD$ means that we have the interaction between a point-like charge and a Dirac point. 

We choose a coordinate system where the Dirac-point is concentrated at position ${{\bf {a}}_{2}}$ with
an magnetic flux $\Phi$. This external source is given by \cite{LHCHLOFAB}
\begin{eqnarray}
\label{dpcharge}
J^{\mu}_{(D)}\left({\bf x}\right)=-2\pi i\Phi\int\frac{d^{3}p}{\left(2\pi\right)^{3}} \ 
\delta\left(p^{0}\right)\epsilon^{0\mu\alpha}p_{\alpha} \ 
e^{-i p\cdot x}e^{-i {\bf{p}}\cdot{\bf {a}_{2}}} \ .
\end{eqnarray}

The expression (\ref{dpcharge}) can be obtained with a dimensional reduction of the field source related to a Dirac string (in $3+1$ dimensions) used in references \cite{Medeiros,DiracString4}. It can be also obtained as a particular case of the source proposed in \cite{KalbRamondFontes} by dimensional reduction.

Substituting (\ref{dpcharge}) in (\ref{corre112Em}), using (\ref{energyS}), discarding self-interacting terms which do not contribute to the force between the Dirac point and the charge (the self-interacting terms are proportional to $\sigma^{2}$ and $\Phi^{2}$), defining the distance vector ${\bf{a}}={\bf {a}}_{1}-{\bf {a}}_{2}$ and following similar steps employed previously, we obtain that
\begin{eqnarray}
\label{Ener10001EM}
E^{CD}=\frac{\sigma\Phi}{m}\left(\int\frac{d^{2}{\bf p}}
{(2\pi)^{2}}e^{i{\bf p}\cdot{\bf a}}-\int\frac{d^{2}{\bf p}}
{(2\pi)^{2}}\frac{{\bf{p}}^{2}}{{\bf p}^2+m^2}e^{{i\bf p}\cdot{\bf a}}\right) \ .
\end{eqnarray}

The first term inside the brackets of Eq. (\ref{Ener10001EM}) is the Dirac delta function $\delta^{2}({\bf{a}})$ and, provided that ${\bf{a}}\not={\bf 0}$, this term vanishes. So, we have
\begin{eqnarray}
\label{Ener100EM}
E^{CD}=\frac{\sigma\Phi}{m}{\bf\nabla}^{2}_{{\bf a}}\int\frac{d^{2}{\bf p}}
{(2\pi)^{2}}\frac{\exp(i{\bf p}\cdot{\bf a})}{{\bf p}^2+m^2} \ ,
\end{eqnarray}
where we defined the differential operator
\begin{eqnarray}
\label{exchange}
{\bf\nabla}_{{\bf a}}=\left(\frac{\partial}{\partial a^{1}},\frac{\partial}{\partial a^{2}}\right) \ .
\end{eqnarray}

Substituting (\ref{int4EM}) in (\ref{Ener100EM}), we arrive at
\begin{eqnarray}
\label{Ener101EM}
E^{CD}= \frac{m\sigma\Phi}{2\pi}K_{0}\left(ma\right) \ .
\end{eqnarray}

The interaction energy (\ref{Ener101EM}) is an effect due to the presence of the term with a Levi-Civita tensor in (\ref{lchho}) (which contains higher order derivatives), and has no counterpart in Maxwell theory, where a point-like charge does not interact with a Dirac point \cite{LHCHLOFAB}. This fact can be verified if one takes into account that the right hand side of (\ref{Ener101EM}) goes to zero when $m$ goes to infinity. 

The corresponding interaction force for (\ref{Ener101EM}) reads
\begin{eqnarray}
\label{For111EM}
F^{CD}=-\frac{d E^{CD}}{d a}=\frac{m^{2}\sigma\Phi}{2\pi}K_{1}(ma) \ .
\end{eqnarray}

We notice that the interaction force (\ref{For111EM}) is repulsive for the case where the charge and the magnetic flux have the same signal, and attractive otherwise. An equivalent situation occurs in Maxwell-Chern-Simons electrodynamics, where the interaction force between the charge and the Dirac point has the same form that the Eq. (\ref{For111EM}) \cite{LHCHLOFAB}. It is due to the fact that the relevant parts of the propagators for this interaction are equal to each other in both theories, what can be seen from
(\ref{propcho}) and (\ref{LDMCS}), as follows 
\begin{eqnarray}
\label{propart2}
D_{CD}^{\mu\nu}\left(p\right)\sim D_{CD(\textrm{MCS})}^{\mu\nu}\left(p\right)\sim -\frac{i}{m\left(p^{2}-m^{2}\right)}\epsilon^{\mu\nu\lambda}p_{\lambda} \ .
\end{eqnarray}
In the limit $m\rightarrow\infty$ the interaction force (\ref{For111EM}) vanishes, as expected.

In the last example, we consider a system composed by two Dirac points. We take a coordinate system where the first Dirac point is placed at the position ${\bf{a}}_{1}$, with magnetic flux $\Phi_{1}$ and the second one, with $\Phi_{2}$, is concentrated at the position ${\bf{a}}_{2}$. This system is described by the external source
\begin{eqnarray}
\label{duasDcurrent1}
J_{\mu}^{DD}\left({\bf x}\right)=J_{\mu(D,1)}\left({\bf x}\right)+J_{\mu(D,2)}\left({\bf x}\right) \ ,
\end{eqnarray}
where
\begin{eqnarray}
\label{duasDcurrent2}
J^{\mu}_{(D,1)}\left({\bf x}\right)=-2\pi i\Phi_{1}\int\frac{d^{3}p}{\left(2\pi\right)^{3}} \ 
\delta\left(p^{0}\right)\epsilon^{0\mu\alpha}p_{\alpha} \ 
e^{-i p\cdot x}e^{-i {\bf{p}}\cdot{\bf {a}_{1}}}\ ,
\end{eqnarray}
and $J^{\mu}_{(D,2)}\left({\bf x}\right)$ is obtained replacing $\Phi$ by $\Phi_{2}$ in (\ref{dpcharge}). The super-index $DD$ means that we have a system composed by two Dirac points.

Substituting (\ref{duasDcurrent2}) in (\ref{energyS}), discarding the self-interacting contributions and proceeding as in the previous cases, we can show that the interaction energy between two Dirac points is given by 
\begin{eqnarray}
\label{Ener10EM}
E^{DD}=-\frac{m^{2}\Phi_{1}\Phi_{2}}{2\pi}K_{0}(ma) \ , 
\end{eqnarray}
which also is similar to the interaction energy between two Dirac points obtained in Ref. \cite{LHCHLOFAB} for the Maxwell-Chern-Simons electrodynamics, but with an overall minus signal. This fact is due to the relevant structure of the propagators (\ref{propcho}) and (\ref{LDMCS}) for this interaction in both theories, which are given by
\begin{eqnarray}
\label{propart3}
D_{DD}^{\mu\nu}\left(p\right)\sim\frac{1}{p^{2}-m^{2}}\eta^{\mu\nu}, \ \ 
D_{DD(\textrm{MCS})}^{\mu\nu}\left(p\right)\sim-\frac{1}{p^{2}-m^{2}}\eta^{\mu\nu} \ .
\end{eqnarray}
Once again, this interaction energy has no counterpart in planar Maxwell electrodynamics \cite{LHCHLOFAB}. 

The interaction force between the Dirac points is given by
\begin{eqnarray}
\label{For2EM}
F^{DD}=-\frac{m^{3}\Phi_{1}\Phi_{2}}{2\pi}K_{1}(m a)  \ .
\end{eqnarray}

The force above is attractive if the two magnetic fluxes have the same signal, and repulsive otherwise. In Maxwell-Chern-Simons electrodynamics an opposite situation is observed \cite{LHCHLOFAB}. It is worth mentioning that the force (\ref{For2EM}) falls down as fast as the distance between the Dirac points, $a$, increases.

As a final comment we point out that by comparing the Eqs. (\ref{For111EM}) and (\ref{For2EM}) with the expression (\ref{forhoch}), we conclude that Dirac points do not behave similarly to point-like charges for the model (\ref{lchho}), on the contrary to the standard Maxwell-Chern-Simons electrodynamics \cite{LHCHLOFAB}.

\section{\label{higherMCS2} The propagator in the presence of a conducting line}

As discussed in Ref. \cite{LHCHLOFAB}, the presence of a conducting line $S$ in the Maxwell-Chern-Simons electrodynamics imposes a boundary condition on the gauge field in such a way that the components of the Lorentz Force parallel on the line vanishes. This condition is attained by
\begin{equation}
\label{condition1}
n^{\mu  \ *}F_{\mu}|_{S} =0 \ ,
\end{equation}
where the sub-index $S$ means that the condition must be taken just on the line $S$, with $ ^{*}F^{\mu}=(1/2)\epsilon^{\mu\nu\lambda}F_{\nu\lambda}$ standing for the dual field strength, and $n^{\mu}$ is the Lorenz three-vector normal to the conducting line. In the higher order derivative extension of the Maxwell-Chern-Simons model, the coupling between the electromagnetic field and charged particles exhibits the same structure as the corresponding one in Maxwell-Chern-Simons electrodynamics. Therefore, the conducting line condition (\ref{condition1}) is the same one for the higher order derivative theory (\ref{lchho}). 

It is important to mention that we have a $2+1$ dimensional model, so a conducting surface is, in fact, just a line. 
 
From now on, we shall consider the presence of a single perfectly conducting line. We shall take a coordinate system where the surface is perpendicular to the $x^{2}$ axis and is located on the line $x^{2}=a$, so that, $n^{\mu}=\eta^{\ \mu}_{2}=\left(0,0,1\right)$ is the Minkowski vector perpendicular to the conducting surface. In this situation, the boundary condition on the gauge field $A^{\mu}$ in (\ref{condition1}) reads
\begin{equation}
\label{condition11}
{^{*}F}_{2}\left(x\right)|_{x^{2}=a}= \epsilon_{2}^{\ \nu\lambda}\partial_{\nu}A_{\lambda}\left(x\right)|_{x^{2}=a}=0 \ .
\end{equation}

By using the functional formalism employed in \cite{LHCHLOFAB,Bordag,LeeWick,LHCFABplate,FABplate}, we can write the functional generator as follows
\begin{eqnarray}
\label{fgen1}
Z_{C}\left[J\right]=\int {\cal{D}}A_{C} \ e^{i\int d^{3}x \ \cal{L}} \ ,
\end{eqnarray}
where the sub-index $C$ means that we are integrating out in all field configurations which satisfy the condition (\ref{condition11}). This restriction is attained by introducing a delta functional, which is non-vanishing only for the field configurations that satisfy the condition (\ref{condition11}), as follows
\begin{eqnarray}
\label{fgen2}
Z_{C}\left[J\right]=\int {\cal{D}}A \ \delta\left[{^{*}F}_{2}\left(x\right)|_{x^{2}=a}\right] \ e^{i\int d^{3}x \ \cal{L}} \ .
\end{eqnarray}

Now we use the Fourier representation for the delta functional
\begin{eqnarray}
\label{fgen3}
\delta\left[{^{*}F}_{2}\left(x\right)|_{x^{2}=a}\right]=\int {\cal{D}}B\exp\left[i
\int d^{3}x\ \delta\left(x^{2}-a\right)B\left(x_{\parallel}\right) {^{*}F_{2}\left(x\right)}\right] \ ,
\end{eqnarray}
where $x_{\parallel}^{\mu}=\left(x^{0},x^{1},0\right)$ means that we have only the coordinates parallel to the conducting surface and $B\left(x_{\parallel}\right)$ is an auxiliary scalar field defined just along the conducting surface and that depends just on the parallel coordinates.

Carrying out similar steps that was employed in reference \cite{LHCHLOFAB}, we can write the functional generator  as follows blue(for more details, see the appendix \ref{AP}).
\begin{eqnarray}
\label{fgen5}
Z_{C}\left[J\right]=Z\left[J\right]{\bar{Z}}\left[J\right] \ ,
\end{eqnarray}
where $Z\left[J\right]$ is the free functional generator (without the conducting surface)
\begin{eqnarray}
\label{fgen6}
Z\left[J\right]=Z\left[0\right]\exp\left[-\frac{i}{2}\int d^{3}x \ d^{3}y \ J^{\mu}\left(x\right)D_{\mu\nu}
\left(x,y\right)J^{\nu}\left(y\right)\right] \ ,
\end{eqnarray}
and ${\bar{Z}}\left[J\right]$ is a contribution due to the scalar field $B$  
\begin{eqnarray}
\label{fgen7}
{\bar{Z}}\left[J\right]=\int{\cal{D}}B\exp\left[i\int d^{3}x \ \delta
\left(x^{2}-a\right)I\left(x\right)B\left(x_{\parallel}\right)\right] \nonumber\\
\times\exp\left[-\frac{i}{2}\int d^{3}x \ d^{3}y \ \delta\left(x^{2}-a\right)
\delta\left(y^{2}-a\right)B\left(x_{\parallel}\right)W\left(x,y\right)
B\left(y_{\parallel}\right)\right] \ ,
\end{eqnarray}
where we identified
\begin{eqnarray}
\label{defi1}
I\left(x\right)=-\int d^{3}y \ \epsilon_{2}^{\ \gamma\alpha}
\left(\frac{\partial}{\partial x^{\gamma}}D_{\alpha\mu}\left(x,y\right)
\right)J^{\mu}\left(y\right) \ , \ W\left(x,y\right)=\epsilon_{2}^{\ \gamma\alpha}
\epsilon_{2}^{\ \beta\lambda}\frac{\partial^{2}D_{\lambda\alpha}\left(x,y\right)}
{\partial x^{\beta}\partial y^{\gamma}} \ .
\end{eqnarray}

Substituting (\ref{defi1}) and (\ref{propcho}) into (\ref{fgen7}), using the fact that \cite{LHCHLOFAB,LeeWick,LHCFABplate} 
\begin{eqnarray}
\label{int}
\int \frac{dp^{2}}{2\pi}\frac{e^{i p^{2}\left(x^{2}-y^{2}\right)}}
{p^{\mu}p_{\mu}-m^{2}}=-\frac{i}{2\Gamma} \ e^{i\Gamma\mid x^{2}-y^{2}\mid} \ , \int \frac{dp^{2}}{2\pi}\frac{e^{i p^{2}\left(x^{2}-y^{2}\right)}}
{p^{\mu}p_{\mu}}=-\frac{i}{2L} \ e^{iL\mid x^{2}-y^{2}\mid} \ ,
\end{eqnarray}
where $p^{2}$ stands for the momentum component perpendicular to the conducing line, $\Gamma=\sqrt{p_{\parallel}^{2}-m^{2}}$ and $L=\sqrt{p_{\parallel}^{2}}$, with the definition of the parallel momentum to the plate $p_{\parallel}^{\mu}=\left(p^{0},p^{1},0\right)$, and defining the parallel metric 
\begin{eqnarray}
\label{etap}
\eta_{\parallel}^{\mu\nu}=\eta^{\mu\nu}-\eta_{\ 2}^{\mu}\eta^{\nu 2} \ ,
\end{eqnarray}
one can write Eq. (\ref{fgen7}) in the following way
\begin{eqnarray}
\label{fgen8}
{\bar{Z}}\left[J\right]={\bar{Z}}\left[0\right]\exp\left[-\frac{i}{2}\int d^{3}x 
\ d^{3}y \ J^{\mu}\left(x\right){\bar{D}}_{\mu\nu}\left(x,y\right)J^{\nu}\left(y\right)\right] \ ,
\end{eqnarray}
where we defined the function (for more details, see the appendix \ref{AP}).
\begin{eqnarray}
\label{proplatehoch}
{\bar{D}}_{\mu\nu}\left(x,y\right)&=&\frac{i}{2}\int \frac{d^{2}p_{\parallel}}
{\left(2\pi\right)^{2}} \ \frac{e^{-i p_{\parallel}\cdot\left(x_{\parallel}
-y_{\parallel}\right)}}{p_{\parallel}^{2}}\frac{1}{\left(\frac{1}{L}-\frac{1}{\Gamma}\right)}\Biggl[\left(\epsilon_{2\gamma\mu}p_{\parallel}^{\gamma}  +\frac{ip_{\parallel}^{2}}{m}\eta_{2\mu}\right)\left(\frac{e^{iL\mid x^{2}-a\mid}}{L}-\frac{e^{i\Gamma\mid x^{2}-a\mid}}{\Gamma}\right)\nonumber\\
&
&-\frac{i}{m}\left(Le^{iL\mid x^{2}-a\mid}-\Gamma e^{i\Gamma\mid x^{2}-a\mid}\right)\eta_{2\mu}\Biggr]\Biggl[\left(\epsilon_{2\beta\nu}p_{\parallel}^{\beta}  -\frac{ip_{\parallel}^{2}}{m}\eta_{2\nu}\right)\left(\frac{e^{iL\mid y^{2}-a\mid}}{L}-\frac{e^{i\Gamma\mid y^{2}-a\mid}}{\Gamma}\right)\nonumber\\
&
&+\frac{i}{m}\left(Le^{iL\mid y^{2}-a\mid}-\Gamma e^{i\Gamma\mid y^{2}-a\mid}\right)\eta_{2\nu}\Biggr] \ .
\end{eqnarray}

Substituting (\ref{fgen8}) and (\ref{fgen6}) in (\ref{fgen5}), the functional generator of the higher order derivative theory (\ref{lchho}) in the presence of a conducting line becomes
\begin{eqnarray}
\label{fgen9}
Z_{C}\left[J\right]=Z_{C}\left[0\right]\exp\left[-\frac{i}{2}
\int d^{3}x \ d^{3}y \ J^{\mu}\left(x\right)\left(D_{\mu\nu}
\left(x,y\right)+{\bar{D}}_{\mu\nu}\left(x,y\right)\right)J^{\nu}
\left(y\right)\right] \ .
\end{eqnarray}

Notice that, from the expression (\ref{fgen9}), one can identify the propagator of the theory in the presence of a conducting line as follows
\begin{eqnarray}
\label{prop3}
D_{C}^{\mu\nu}=D^{\mu\nu}\left(x,y\right)+{\bar{D}}^{\mu\nu}\left(x,y\right) \ .
\end{eqnarray}

The propagator (\ref{prop3}) is composed by the sum of the free propagator (\ref{propcho}) with the correction (\ref{proplatehoch}), which accounts for the presence of the conducting linear surface. It can be checked out that taking the limit $m\rightarrow\infty$ in (\ref{proplatehoch}) we recover the standard Maxwell propagator in the presence of a conducting line (in $3d$), and that the conducting line condition (\ref{condition11}) is really satisfied. Besides, we can also show that
\begin{eqnarray}
\label{conditionhomcs}
{\cal{O}}_{\mu\nu}D^{\nu\alpha}_{C}\left(x,y\right)=\eta_{\mu}^{\ \alpha}\delta^{3}\left(x-y\right) \ ,
\end{eqnarray}
where ${\cal{O}}_{\mu\nu}$ is the operator defined in (\ref{diffopehomcs}), what means that the gauge field propagator under the boundary conditions (\ref{condition11}) is really a Green function for the problem.

\section{\label{higherMCS3} Particle-Conductor interaction}

In this section we consider the interaction between a point-like charge and the conducting line. We can show that the interaction energy between a static source $J^{\mu}\left(x\right)$ and a conducting surface, in a quadratic theory, is given by \cite{LHCHLOFAB,LeeWick,LHCFABplate,FAFEB}
\begin{eqnarray}
\label{energy}
{{E}}=\frac{1}{2T}\int d^{3}x \ d^{3}y \ J^{\mu}\left(x\right)
{\bar{D}}_{\mu\nu}\left(x,y\right)J^{\nu}\left(y\right) \ .
\end{eqnarray}

The presence of a point-like charge is accomplished by the external source
\begin{eqnarray}
\label{source1}
J^{C}_{\mu}\left(x\right)=q\eta^{0}_{\ \mu}\delta^{2}\left({\bf x}-{\bf b}\right) \ ,
\end{eqnarray}
where ${\bf{b}}$ is a constant vector standing for the charge position that will be taken to be ${\bf b} = (0, b)$, from now on, for the sake of simplicity. 

Substituting (\ref{source1}) and (\ref{proplatehoch}) in (\ref{energy}), and then performing some manipulations similar to the ones employed in Sect. \ref{higherMCS1b}, we obtain
\begin{eqnarray}
\label{enerppar}
E^{LC}=-\frac{q^{2}}{4\pi}\int^{\infty}_{0}d|{\bf{p}}_{\parallel}|\frac{\sqrt{{\bf{p}}_{\parallel}^{2}}\sqrt{{\bf{p}}_{\parallel}^{2}+m^{2}}}{\sqrt{{\bf{p}}_{\parallel}^{2}+m^{2}}-\sqrt{{\bf{p}}_{\parallel}^{2}}}\left(\frac{e^{-R\sqrt{{\bf{p}}_{\parallel}^{2}}}}{\sqrt{{\bf{p}}_{\parallel}^{2}}}-\frac{e^{-R\sqrt{{\bf{p}}_{\parallel}^{2}+m^{2}}}}{\sqrt{{\bf{p}}_{\parallel}^{2}+m^{2}}}\right)^{2} \ ,
\end{eqnarray}
where $R=\mid b-a\mid$ stands for the distance between the plate and the charge and the super-index $LC$ means that we have the interaction energy between the conducting line and the charge.

Equation (\ref{enerppar}) can be simplified with the change of integration variable $p =\mid{\bf{p}}_{\parallel}\mid/m$,
\begin{eqnarray}
\label{cplatehoc}
E^{LC}&=&-\frac{q^{2}}{4\pi}\int_{0}^{\infty} dp \ p\left[\left(p^{2}+1\right)+p\sqrt{p^{2}+1}\right]\Biggl(\frac{e^{-2pmR}}{p^{2}}-2\frac{e^{-\left(p+\sqrt{p^{2}+1}\right)mR}}{p\sqrt{p^{2}+1}}+\frac{e^{-2mR\sqrt{p^{2}+1}}}{p^{2}+1}\Biggr) \ .
\end{eqnarray}

Each contribution in the integral (\ref{cplatehoc}) can be calculated exactly. For the first contribution, we have
\begin{eqnarray}
\label{contribution1}
\int_{0}^{\infty} dp \left[\left(p^{2}+1\right)+p\sqrt{p^{2}+1}\right]\frac{e^{-2pmR}}{p}=\frac{1}{4\left(mR\right)^{2}}+\frac{\pi}{4mR}\left[SH_{1}\left(2mR\right)-Y_{1}\left(2mR\right)\right]+\int_{0}^{\infty} dp \frac{e^{-2pmR}}{p} \ ,
\end{eqnarray}
where $Y$ and $SH$ stand for the Bessel function of second kind and the Struve function, respectively \cite{Arfken}. We notice that the integral on the right-hand side of the Eq. (\ref{contribution1}) is divergent. It can be regularized by inserting a parameter $\epsilon$, as follows
\begin{eqnarray}
\label{contribution12}
\int_{0}^{\infty} dp \frac{e^{-2pmR}}{p}=\lim_{\epsilon\rightarrow 0}\int_{\epsilon}^{\infty} dp \frac{e^{-2pmR}}{p}=\lim_{\epsilon\rightarrow 0}\left[Ei\left(1,2mR\epsilon\right)\right] \ ,
\end{eqnarray}
where the limit $\epsilon\to0$ is taken from the right due to the definition of the defined integral and $Ei\left(n, s\right)$ is the exponential integral function \cite{Arfken}, defined by 
\begin{equation}
\label{Ei}
Ei\left(n, s\right)=\int^{\infty}_{1}\frac{e^{-ts}}{t^{n}}dt \ \ \ {\Re}\left(s\right)>0 \ , \ n = 0, 1, 2, \cdots \ .
\end{equation}

It is worth mentioning that $Ei(1,x)=\Gamma(0,x)$, for $x\in\Re$.

With the aid the approximation for $Ei\left(1,2mR\epsilon\right)$ for small arguments \cite{Arfken}, one can write 
\begin{equation}
\label{contribution13}
Ei\left(1,2mR\epsilon\right)\stackrel{\epsilon\rightarrow0^{+}}{\rightarrow}-\gamma-\ln\left(2mR\right)-\ln\epsilon+{\cal O}(\epsilon) \ . 
\end{equation}
Therefore,
\begin{eqnarray}
\label{intregulari}
\int_{0}^{\infty} dp \frac{e^{-2pmR}}{p}&=&\lim_{\epsilon\rightarrow 0}\left[-\ln\left(2mR\right)-\gamma-\ln\epsilon\right]\nonumber\\
&=&\lim_{\epsilon\rightarrow 0}\left[-\ln\left(2mR\right)-\gamma-\ln\epsilon+\ln\left(2mR_{0}\right)-\ln\left(2mR_{0}\right)\right]\nonumber\\
&=&-\ln\left(\frac{R}{R_{0}}\right)-\gamma-\lim_{\epsilon\rightarrow 0}\ln\left(2mR_{0}\epsilon\right)\nonumber\\
&\rightarrow &-\ln\left(\frac{R}{R_{0}}\right) \ ,
\end{eqnarray}
where $\gamma$ is the Euler constant and $R_{0}$ is an arbitrary constant with dimension of length. In the last line of eq. (\ref{intregulari}) we neglected a divergent term that does not depend on the distance $R$, once it does not contribute to the interaction force between the charge and the conducting line. The arbitrary constant $R_{0}$ does not have any special physical meaning and does not contribute to the force between the charge and the conductor. It was just introduced simply to make the argument of the $\ln$ function dimensionless and characterizes a redefinition of the zero of the interaction energy.

For the third contribution on the right hand side of (\ref{cplatehoc}) we perform the change in the integration variable $u=\sqrt{p^{2}+1}$, as follows
\begin{eqnarray}
\label{contribution2}
\int_{0}^{\infty} dp\ p\left[1+p\left(p^{2}+1\right)^{-1/2}\right]e^{-2mR\sqrt{p^{2}+1}}&=&\int_{1}^{\infty}du\left(u+\sqrt{u^{2}-1}\right)e^{-2umR}\nonumber\\
&=& \frac{K_{1}\left(2mR\right)}{2mR}+\frac{e^{-2mR}}{4\left(mR\right)^{2}}\left(1+2mR\right) \ .
\end{eqnarray}

The second contribution to (\ref{cplatehoc}) is obtained with the change of variable $u=p+\sqrt{p^{2}+1}$, as follows
\begin{eqnarray}
\label{contribution3}
-2\int_{0}^{\infty} dp\left(p+\sqrt{p^{2}+1}\right)e^{-\left(p+\sqrt{p^{2}+1}\right)mR}&=&-\int_{1}^{\infty}du\frac{u^{2}+1}{u}e^{-mRu}\nonumber\\
&=&-e^{-mR}\left(\frac{1}{\left(mR\right)}+\frac{1}{\left(mR\right)^{2}}\right)-Ei\left(1,mR\right)\ .
\end{eqnarray}

Putting all this together, we have the interaction energy between the point-charge and the conducting line
\begin{eqnarray}
\label{cplate2}
E^{LC}&=&-\frac{q^{2}}{4\pi}\left[-\ln\left(\frac{R}{R_{0}}\right)+\Delta_{1}\left(mR\right)\right] \ ,
\end{eqnarray}
where we defined the function,
\begin{eqnarray}
\label{delta1mr}
\Delta_{1}\left(mR\right)&=&\frac{1}{4\left(mR\right)^{2}}+\frac{\pi}{4mR}\left[SH_{1}\left(2mR\right)-Y_{1}\left(2mR\right)\right]-e^{-mR}\left(\frac{1}{\left(mR\right)}+\frac{1}{\left(mR\right)^{2}}\right)\nonumber\\
&
&-Ei\left(1,mR\right)+\frac{K_{1}\left(2mR\right)}{2mR}+\frac{e^{-2mR}}{4\left(mR\right)^{2}}\left(1+2mR\right) \ .
\end{eqnarray}

The result (\ref{cplate2}) is exact, but difficult to be interpreted. The first term on the right hand side is the same as the one found for the surface-charge interaction obtained in standard $3d$ Maxwell Electrodynamics. It is important to mention that the Coulomb energy in two space dimensions exhibits a logarithmic behavior. The second term falls when $mR$ increases faster than the first term. 

From the Eq. (\ref{cplate2}) we obtain interaction force between the conducting line and the charge
\begin{eqnarray}
\label{FFCP}
F^{LC}=-\frac{q^{2}}{4\pi R}\left[1+\Delta_{2}\left(mR\right)\right] \ ,
\end{eqnarray}
where the function $\Delta_{2}\left(mR\right)$ is defined by
\begin{eqnarray}
\label{deltamr}
\Delta_{2}\left(mR\right)&=&\frac{1}{2\left(mR\right)^{2}}-\frac{\pi}{2}\left[Y_{2}\left(2mR\right)+SH_{0}\left(2mR\right)-\frac{SH_{1}\left(2mR\right)}{\left(mR\right)}\right]+K_{2}\left(2mR\right)\nonumber\\
&
&-2e^{-mR}\left[1+\frac{1}{\left(mR\right)}+\frac{1}{\left(mR\right)^{2}}\right]+e^{-2mR}\left[1+\frac{1}{\left(mR\right)}+\frac{1}{2\left(mR\right)^{2}}\right] \ .
\end{eqnarray}

\begin{figure}[!h]
\centering \includegraphics[scale=0.45]{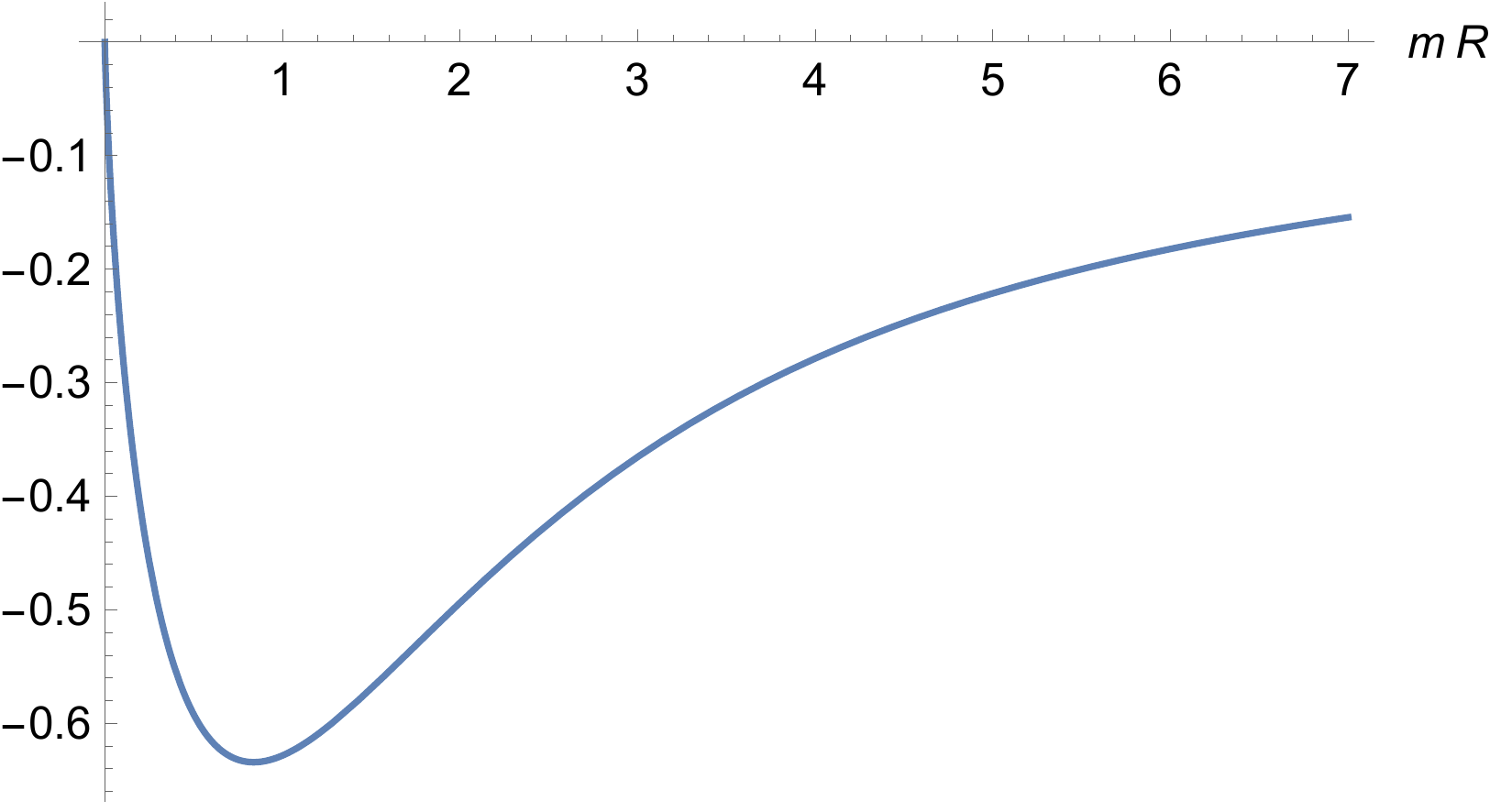} \caption{Plot for $\frac{4\pi F^{LC}}{mq^{2}}$, from (\ref{FFCP}), as a function of $mR$.}
\label{LinhaCarga} 
\end{figure}

The first term on the right hand side of the Eq. (\ref{FFCP}) is the usual Coulomb interaction (in $3d$) between the charge $q$ and its image, placed at a distance $2R$ apart. The second term is a correction imposed by the parameter $m$, which falls down when $mR$ increases. We notice that the interaction force (\ref{FFCP}) is always attractive, since the term inside brackets on the right hand side is always positive. In Fig. (\ref{LinhaCarga}) we have a plot for the force (\ref{FFCP}) multiplied by $\frac{4\pi}{mq^{2}}$ as a function of $mR$. We can see that there is a global minimum around $mR\cong 0.82$ and just one zero in the limit $mR=0$, when the charge approaches to the conducting line.

The reader could ask what is the true value of the force (\ref{FFCP}) when $mR=0$, once the results obtained up to now were taken on the assumption that $mR>0$. In this situation the charge is really taken to be lying on the conducting line. To answer this question we have to go back to expression (\ref{cplatehoc}), calculate the force and take case $mR=0$, as follows
\begin{eqnarray}
F^{LC}(mR=0)=-\frac{\partial E^{LC}}{\partial R}|_{mR=0}&=&-\frac{q^{2}m}{2\pi}\int_{0}^{\infty} dp \ p\left[\left(p^{2}+1\right)+p\sqrt{p^{2}+1}\right]\Bigl(\frac{e^{-2pmR}}{p}\cr
&\ &-(p+\sqrt{p^{2}+1})\frac{e^{-\left(p+\sqrt{p^{2}+1}\right)mR}}{p\sqrt{p^{2}+1}}+\frac{e^{-2mR\sqrt{p^{2}+1}}}{\sqrt{p^{2}+1}}\Bigr)\Big|_{mR=0}\ .
\end{eqnarray}
With some simple manipulations, one can show that the integrand above vanishes and so, also the force for $mR=0$.

In order to check the validity, or not, of the image method, we consider the expression (\ref{forhoch}) for the special case where $\sigma_{1}=q$, $\sigma_{2}=-q$ and $a=2R$, 
\begin{equation}
\label{fimv}
F^{CC}= -\frac{q^{2}}{4\pi R}\left[1-(2mR)K_{1}\left(2mR\right)\right] \ .
\end{equation}

We notice that Eq. (\ref{fimv}) is different from Eq. (\ref{FFCP}) thus, on the contrary to the Maxwell-Chern-Simons theory \cite{LHCHLOFAB}, the image method is not valid for the model (\ref{lchho}) with the conducting line condition (\ref{condition11}). A similar situation occurs in the $4d$ Lee-Wick electrodynamics with the presence of a conducting plate, where the image method is not valid \cite{LeeWick}. For the $3d$ Lee-Wick electrodynamics with the presence of a conducting line, we hope a similar situation \cite{progress}.

The non-validity of the image method in the $4d$ Lee-Wick electrodynamics is related to the non-triviality of the boundary conditions imposed by conductors in this theory, because we have two kinds of field modes in this case, some of them being  massless and other having mass. It is evinced in the formulation of the Lee-Wick electrodynamics in terms of two fields \cite{Accioly}. Maybe it is an indication that the $3d$ model (\ref{lchho}) could be written in terms of two coupled fields, similarly to the Lee-Wick electrodynamics.

Besides, it is important to point out that the image method is based on the fact that, in Maxwell electrostatics, the field configurations can be obtained  with only the zero component of the gauge field (in an appropriated gauge), which in turn obeys the Poisson equation and directly gives the energy of the system. For the model (\ref{lchho}) with higher order derivatives, even for stationary situations, the zero component of the potential does not obey the Poisson equation, but instead, it is a solution of an equation with higher order derivatives. Furthermore, for models with higher order derivatives, energies of stationary systems are not obtained directly from the zero component of the gauge field. As an example, one can see the 00 component of the energy momentum tensor of the Lee Wick electrodynamics \cite{CasimirLeeWick}.

The force (\ref{fimv}) falls down when $mR$ increases and is always attractive, similarly to (\ref{FFCP}). In the limit where $R\rightarrow 0$, both forces (\ref{FFCP}) and (\ref{fimv}) are not divergent, but go to zero. This fact is due to the presence of the higher order derivatives term in the model (\ref{lchho}). It is a new example where the presence of higher order derivatives (in this case, in a term with the Levi-Civita tensor) can improve renormalization properties and tame ultraviolet divergences \cite{autoenergiaLW}, even with the presence of material boundaries \cite{LeeWick}.

The deviation from the image method behavior can be seen from the difference between Eqs. (\ref{FFCP}) and (\ref{fimv}) normalized by the Coulombian force in $2+1$ dimensions, as follows,
\begin{equation}
\label{deviation}
\delta\left(mR\right)=\frac{\mid F^{LC}\mid-\mid F^{CC}\mid}{\left[q^{2}/\left(4\pi R\right)\right]}=\Delta_{2}\left(mR\right)+(2mR)K_{1}\left(2mR\right) \ .
\end{equation}

\begin{figure}[!h]
\centering \includegraphics[scale=0.50]{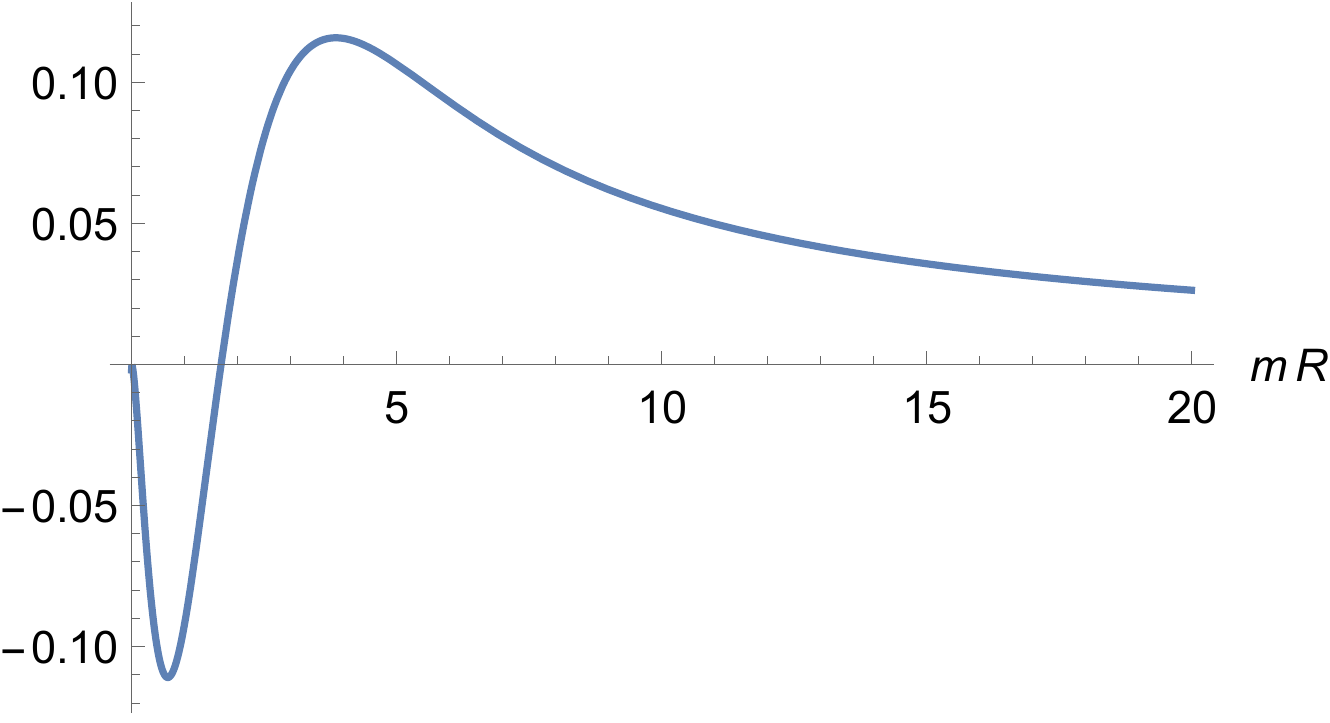} \caption{Plot for $\delta\left(mR\right)$.}
\label{graficocalF|} 
\end{figure}

In Fig. (\ref{graficocalF|}) we have a plot for $\delta\left(mR\right)$ as a function of $mR$. In the limit $mR\rightarrow\infty$ we have $\delta\rightarrow 0$. In the interval $0 < mR\approx< 1,72$ we have $\delta<0$ and the modulus of charge-line interaction is smaller than the modulus of charge-image interaction. For $mR >\approx 1,72$ we have $\delta > 0$ and the charge-line interaction is greater than the charge-image interaction, in modulus. It is also interesting to notice that the curve of Fig. (\ref{graficocalF|}) exhibits a maximum for $mR\cong 3,82$, a minimum for $mR\cong 0,72$ and two zeros for $mR=0$ and $mR\cong 1,72$.

\section{\label{higherMCS4} Dirac point-Conductor interaction}

In this section we study the interaction between a Dirac point and the conducting line. This kind of interaction does not occur in Maxwell electrodynamics \cite{LHCHLOFAB}.

First, we consider de Dirac point placed at the position ${\bf{b}}=(0,b)$, as follows
\begin{eqnarray}
\label{dpplate}
J^{\mu}\left({\bf x}\right)=-2\pi i\Phi\int\frac{d^{3}p}{\left(2\pi\right)^{3}} \ 
\delta\left(p^{0}\right)\epsilon^{0\mu\alpha}p_{\alpha} \ 
e^{-i p\cdot x}e^{-i {\bf{p}}\cdot{{\bf{b}}}} \ .
\end{eqnarray}

Substituting (\ref{dpplate}) and (\ref{proplatehoch}) in (\ref{energy}) and following the same steps employed in the previous section, we obtain
\begin{eqnarray}
\label{dpplate22}
E^{LD}=-\frac{m^{2}\Phi^{2}}{4\pi}\int_{0}^{\infty} dp\ p\left(\sqrt{p^{2}+1}+p\right)\frac{e^{-2mR\sqrt{p^{2}+1}}}{\sqrt{p^{2}+1}} \ ,
\end{eqnarray}
where the superscript $LD$ means that we have the interaction between the Dirac point and the conducting line and $R=\mid b-a\mid$ stands for the distance between the conductor and the Dirac point.

Now, by using the Eq. (\ref{contribution2}), we arrive at
\begin{eqnarray}
\label{dpplate2}
E^{LD}=-\frac{m^{2}\Phi^{2}}{4\pi}\left[\frac{K_{1}\left(2mR\right)}{2mR}+\frac{e^{-2mR}}{4\left(mR\right)^{2}}\left(1+2mR\right)\right] \ .
\end{eqnarray}

The interaction energy (\ref{dpplate2}) falls down when $mR$ increases and vanishes in the limit $m\rightarrow\infty$. 

The interaction force reads
\begin{eqnarray}
\label{fdpplate}
F^{LD}=-\frac{m^{2}\Phi^{2}}{4\pi R}\left[e^{-2mR}\left(1+\frac{1}{\left(mR\right)}+\frac{1}{2\left(mR\right)^{2}}\right)+K_{2}\left(2mR\right)\right] \ ,
\end{eqnarray}
which is always attractive. 

\begin{figure}[!h]
\centering \includegraphics[scale=0.45]{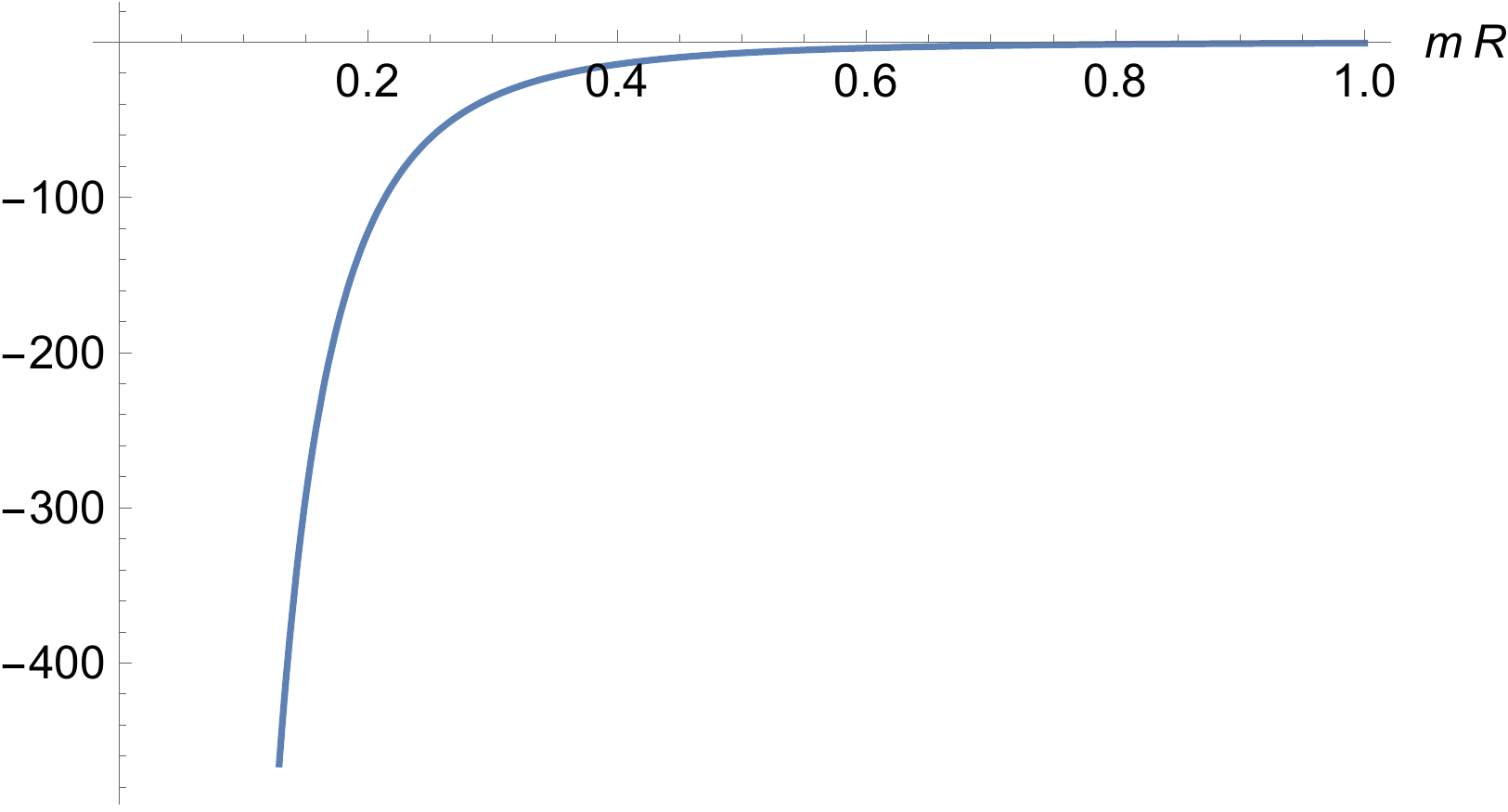} \caption{Plot for $\frac{4\pi F^{LD}}{m^3\phi^{2}}$, from (\ref{fdpplate}), as a function of $mR$.}
\label{LinhaDirac} 
\end{figure}

The interaction force between the surface and the Dirac point diverges when the source is placed on the conducting line. In Fig. (\ref{LinhaDirac}) we have a plot for the force (\ref{fdpplate}) multiplied by $\frac{4\pi}{m^3\phi^{2}}$ as a function of $mR$.

It can be checked that for the case where $\Phi_{1}=-\Phi_{2}=\Phi$, $a=2R$, the Eq. (\ref{For2EM}) turns out to be different from the expression (\ref{fdpplate}). So, the image method is not valid for the Dirac point in the presence of the conducting line condition (\ref{condition11}). This situation is different to the one found in the Maxwell-Chern-Simons electrodynamics, where the image method remains valid for the Dirac point \cite{LHCHLOFAB}.

\section{Conclusions}
\label{conclusoes}

In this paper, we have investigated the interactions between stationary point-like sources for the higher order derivatives extension of $3d$ Chern-Simons model. We have considered effects related to field sources which describe point-like charges and Dirac points. 

Afterwards, we have considered the same model with the presence of a perfectly conducting linear surface (notice that we have a model defined in a $(2+1)$-dimensional spacetime.) The propagator for the gauge field in the presence of the conducting surface have been calculated exactly. We have obtained the interaction force between the conductor and a point-like charge as well as the force between the conductor and a Dirac point. It have been shown that the image method is not valid for the model (\ref{lchho}) for any setup that we have been considered. For the interaction between the point-like charge and the conductor, we have a local minimum for a given value for the distance.

The non validity of the image method for the model (\ref{lchho}) is an indication that, maybe, the model (\ref{lchho}) could be written in terms of two coupled fields, as the $4d$ Lee-Wick electrodynamics.

We have compared the results obtained along the paper with the corresponding ones obtained from the Maxwell-Chern-Simons electrodynamics \cite{LHCHLOFAB}.

{\bf Acknowledgments:}\\
For financial support, L.H.C. Borges thank to CAPES (Brazilian agency) and FAPESP (Grant 2016/11137-5), F.A. Barone thanks to CNPq (Brazilian agency) under the grant 313978/2018-2 and H.L. Oliveira thanks to CAPES (Brazilian agency).

\appendix

\section{The Propagator}
\label{AP}

In this appendix we give some additional technical details of how the propagator in Eq. (\ref{proplatehoch}) was       computed. We start by substituting  (\ref{fgen3}) in (\ref{fgen2}) and using Eq. (\ref{condition11}), what leads to
\begin{eqnarray}
\label{fgen46}
Z_{C}\left[J\right]=\int {\cal{D}}A{\cal{D}}B\ e^{i\int d^{3}x \ \cal{L}}
\exp\left[-i\int d^{3}x\ \delta\left(x^{2}-a\right)A_{\beta}\left(x\right)
\epsilon_{2}^{\ \alpha\beta}\partial_{\alpha}B\left(x_{\parallel}\right)\right] \ .
\end{eqnarray}

We can see that the first exponential in (\ref{fgen46}) depends only on the gauge field $A^{\mu}$, but the second one involves a coupling between the fields $A^{\mu}$ and $B$. In order to decouple $A^{\mu}$ and $B$, we carry out the following translation
\begin{eqnarray}
\label{trans1}
A^{\beta}\left(x\right)\rightarrow A^{\beta}\left(x\right)+\int d^{3}y\ 
D^{\beta}_{\ \alpha}\left(x,y\right)\delta\left(y^{2}-a\right)\epsilon_{2}
^{\ \gamma\alpha}\partial_{\gamma}B\left(y_{\parallel}\right) \ ,
\end{eqnarray}
which has an unitary jacobian and allows us to write the functional generator (\ref{fgen46}) in the form (\ref{fgen5}).

Substituting (\ref{defi1}) into (\ref{fgen7}) and using the Eqs. (\ref{propcho}), (\ref{int}) and (\ref{etap}), we arrive at
\begin{eqnarray}
\label{FJP}
{\bar{Z}}\left[J\right]=\int{\cal{D}}B\exp\left[i\int d^{2}x_{\parallel} \ I\left(x_{\parallel}\right)B\left(x_{\parallel}\right)\right]\exp\left[-\frac{i}{2}\int d^{2}x_{\parallel} \ d^{2}y_{\parallel} \ B\left(x_{\parallel}\right)W\left(x_{\parallel},y_{\parallel}\right)
B\left(y_{\parallel}\right)\right] \ ,
\end{eqnarray}
where 
\begin{eqnarray}
\label{WIP}
W\left(x_{\parallel},y_{\parallel}\right)&=&-\frac{i}{2}\int\frac{d^{2}p_{\parallel}}{\left(2\pi\right)^{2}}e^{-ip_{\parallel}\cdot\left(x_{\parallel}-y_{\parallel}\right)}p_{\parallel}^{2}\left(\frac{1}{L}-\frac{1}{\Gamma}\right)\ ,\cr
I\left(x_{\parallel}\right)&=&\int d^{3}y\ f_{\mu}\left(y,x_{\parallel}\right)J^{\mu}\left(y\right) \ ,
\end{eqnarray}  
with the definition
\begin{eqnarray}
\label{WIP2}
f_{\mu}\left(y,x_{\parallel}\right)&=&-\frac{1}{2}\int\frac{d^{2}p_{\parallel}}{\left(2\pi\right)^{2}}e^{-ip_{\parallel}\cdot\left(x_{\parallel}-y_{\parallel}\right)}\Biggl[\left(\epsilon_{2\gamma\mu}p_{\parallel}^{\gamma}-\frac{ip_{\parallel}^{2}}{m}\eta_{2\mu}\right)\left(\frac{e^{iL\mid y^{2}-a\mid}}{L}-\frac{e^{i\Gamma\mid y^{2}-a\mid}}{\Gamma}\right)\nonumber\\
&
&+\frac{i}{m}\left(Le^{iL\mid y^{2}-a\mid}-\Gamma e^{i\Gamma\mid y^{2}-a\mid}\right)\eta_{2\mu}\Biggr] \ .
\end{eqnarray}

Now, in the functional integral (\ref{FJP}) we perform the following translation,
\begin{eqnarray}
\label{tran2}
B\left(x_{\parallel}\right)\rightarrow 
B\left(x_{\parallel}\right)+\int d^{2}y_{\parallel} \ 
V\left(x_{\parallel},y_{\parallel}\right)I\left(y_{\parallel}\right) \ ,
\end{eqnarray}
where $V\left(x_{\parallel},y_{\parallel}\right)$ is the function which inverts $W\left(x_{\parallel},y_{\parallel}\right)$, in the sense that,
\begin{eqnarray}
\label{VF2}
\int d^{2} y_{\parallel}W\left(x_{\parallel},y_{\parallel}\right)V\left(y_{\parallel},z_{\parallel}\right)=\delta^{2}\left(x_{\parallel}-z_{\parallel}\right) \ ,
\end{eqnarray}
namely,
\begin{eqnarray}
\label{VF1}
V\left(x_{\parallel},y_{\parallel}\right)=2i\int\frac{d^{2}p_{\parallel}}{\left(2\pi\right)^{2}}e^{-ip_{\parallel}\cdot\left(x_{\parallel}-y_{\parallel}\right)}\frac{1}{p_{\parallel}^{2}\left(\frac{1}{L}-\frac{1}{\Gamma}\right)} \ .
\end{eqnarray}

With the translation (\ref{tran2}) we obtain the functional generator (\ref{fgen8}), where the correction to the propagator which accounts for the  presence of the conducting line is given by 
\begin{eqnarray}
\label{CDPROP}
{\bar{D}}_{\mu\nu}\left(x,y\right)=-\int d^{2} z_{\parallel} \ d^{2}w_{\parallel} \ f_{\mu}\left(x,z_{\parallel}\right)V\left(z_{\parallel},w_{\parallel}\right)f_{\nu}\left(y,w_{\parallel}\right) \ .
\end{eqnarray}

Substituting (\ref{WIP2}) and (\ref{VF1}) into Eq. (\ref{CDPROP}) and performing some calculations, we finally obtain the expression (\ref{proplatehoch}).

\end{document}